\documentclass[10pt]{article}
\usepackage{graphicx}
\usepackage{caption}
\usepackage{subcaption}

\def\met{\mbox{${\hbox{$E$\kern-0.65em\lower-.1ex\hbox{\bf{/}}}}_T$}} 
\def\mex{\mbox{${\hbox{$E$\kern-0.65em\lower-.1ex\hbox{\bf{/}}}}_x$}~} 
\def\mey{\mbox{${\hbox{$E$\kern-0.65em\lower-.1ex\hbox{\bf{/}}}}_y$}~} 

\def\Title#1{\begin{center} {\Large #1 } \end{center}}
\def\Author#1{\begin{center}{ \sc #1} \end{center}}
\def\Address#1{\begin{center}{ \it #1} \end{center}}

\newcommand\pubblock{\rightline{\begin{tabular}{l} Proceedings of the Second Annual LHCP\\ \pubnumber\\
         \pubdate  \end{tabular}}}

\newenvironment{Abstract}{\begin{quotation} \begin{center} 
             \large ABSTRACT \end{center}\bigskip 
      \begin{center}\begin{large}}{\end{large}\end{center} \end{quotation}}

\newenvironment{Presented}{\begin{quotation} \begin{center} 
             PRESENTED AT\end{center}\bigskip 
      \begin{center}\begin{large}}{\end{large}\end{center} \end{quotation}}

\def\beq{\begin{equation}}
\def\eeq#1{\label{#1}\end{equation}}
\def\eeqn{\end{equation}}

\def\beqa{\begin{eqnarray}}
\def\eeqa#1{\label{#1}\end{eqnarray}}
\def\eeqan{\end{eqnarray}}

\let\bar=\overbar

\def\Dslash{\not{\hbox{\kern-4pt $D$}}}
\def\dslash{\not{\hbox{\kern-2pt $\del$}}}

\def\msb{{\bar{\ssstyle M \kern -1pt S}}}

\usepackage{color}

 \newcommand\pubnumber{ CMS EXPERIMENT}

\newcommand\pubdate{\today}

\def\affiliation{
On behalf of the CMS Experiment, \\
Department of Physics \\
Brown University, Providence, RI 02912, U.S.A }

\begin{document}

\large
\begin{titlepage}
\pubblock

\vfill
\Title{  Performance of e/$\gamma$-based Triggers at the CMS High Level Trigger }
\vfill

\Author{ Zeynep Demiragli }
\Address{\affiliation}
\vfill
\begin{Abstract}

The CMS experiment has been designed with a two-level trigger system: the Level 1 (L1) Trigger, implemented on custom-designed electronics, and the High Level Trigger (HLT), a streamlined version of the CMS reconstruction and analysis software running on a computer farm. In order to achieve a good rate reduction with as little as possible impact on the physics efficiency, the algorithms used at HLT are designed to follow as closely as possible the ones used in the offline reconstruction. Here, we will present the algorithms used for the online reconstruction of electrons and photons (e/$\gamma$), both at L1 and HLT, and their performance and the planned improvements of these HLT objects.
\end{Abstract}
\vfill

\begin{Presented}
The Second Annual Conference\\
 on Large Hadron Collider Physics \\
Columbia University, New York, U.S.A \\ 
June 2-7, 2014
\end{Presented}
\vfill
\end{titlepage}
\def\thefootnote{\fnsymbol{footnote}}
\setcounter{footnote}{0}
%

\normalsize 


\section{Introduction}

The electromagnetic calorimeter (ECAL) of CMS give an accurate measurement of energy and position of electrons and photons (e/$\gamma$) used in online and offline analysis. Here, we will present the algorithms used for the online reconstruction of e/$\gamma$, both at L1 and HLT, and their performance.

\section{Level-1 Trigger Algorithms and Performance}

\begin{figure}[htb]
\centering
\begin{subfigure}{0.48\textwidth}
\centering
\includegraphics[height=2.3in]{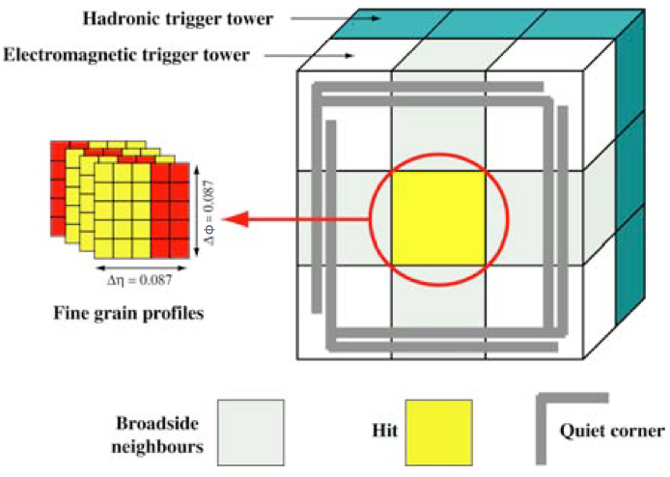}
\caption{Pictorial representation of the transverse energy calculation of an e/$\gamma$ deposition}
\end{subfigure}
$\;\;\;\;$
\begin{subfigure}{0.48\textwidth}
\centering
\includegraphics[height=2.3in]{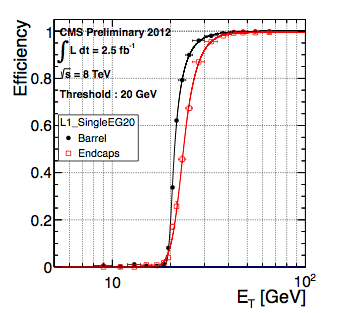}
\caption{Efficiency of the 20 GeV e/$\gamma$ trigger as a function of offline $E_{T}$, shown for EB and EE.}
\end{subfigure}
\caption{} \label{fig:L1}
\end{figure}

The e/$\gamma$ Level-1 trigger algorithm uses a sliding window of 3 $\times$ 3 trigger towers (TTs). The transverse energy of the e/$\gamma$ candidate is given by the energy deposit of the central tower summed with the largest deposit in one of its 4 neighbour towers as shown in Fig.~\ref{fig:L1}a. Only candidates in which 2 adjacent strips (5 crystals in $\phi$ are called a strip) of the central tower contain 90\% of the tower energy are kept. The associated HCAL energy contribution is required to be less than 5\% of the ECAL energy. To qualify as an isolated candidate, at least 5 of the 8 adjacent TTs must have transverse energy below a certain threshold.

The L1 trigger efficiency is measured with Z(ee) events using a tag and probe method. Both the tag and the probe are required to pass tight identification cuts and the tag must also trigger the event at L1. The trigger efficiency is defined as the fraction of probes matched to a L1 candidate, and is given as a function of the probe ET. The widths of the turn-on curves are affected by the coarse trigger granularity, which degrade the energy resolution at L1. In the ECAL endcap (EE) region, the amount of material in front of the calorimeter is larger increasing the probability of bremstrahlung emission. The turn-on curve in EE results to be wider than in barrel. (EB). An example is shown in Fig.~\ref{fig:L1}b.

\section{HLT Algorithms and Performance}

\begin{figure}[htb]
\centering
\begin{subfigure}{0.48\textwidth}
\centering
\includegraphics[height=2.3in]{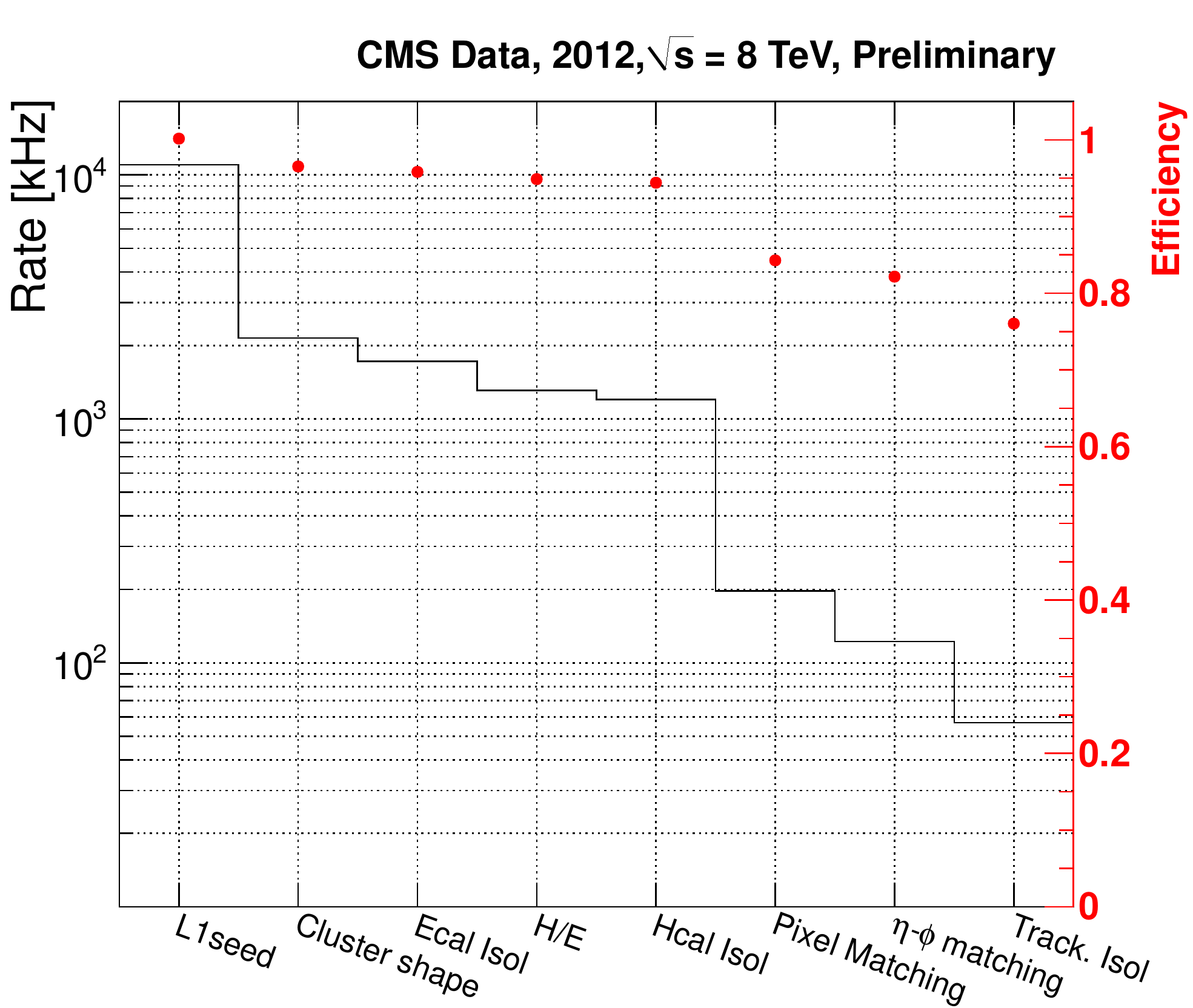}
\caption{The plot shows how the rate is gradually reduced by the filtering steps of this trigger (black) along with the efficiency on electrons (red)}
\end{subfigure}
$\;\;\;\;$
\begin{subfigure}{0.48\textwidth}
\centering
\includegraphics[height=2.3in]{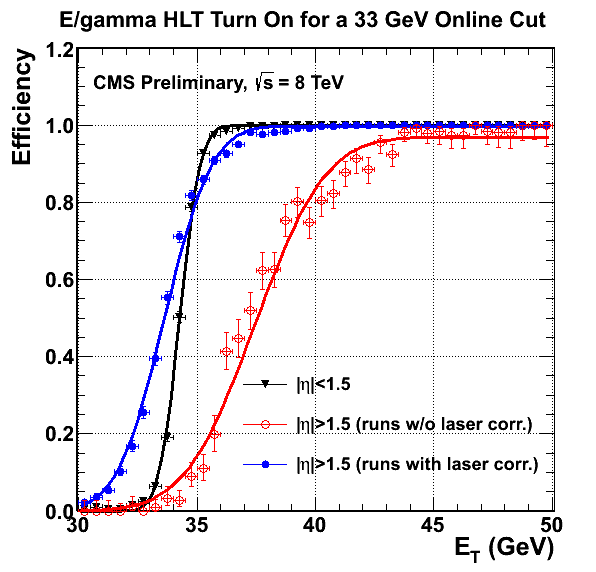}
\caption{Electron efficiency as a function of offline ET  in the EB and EE before and after the transparency corrections at HLT}
\end{subfigure}
\caption{} \label{fig:ele}
\end{figure}

HLT algorithms for e/$\gamma$ selection begin by combining energy deposits around a L1 candidate to form Super Clusters (SC) using the same algorithms as in the offline reconstruction. Clustering algorithms account for the spread of bremsstrahlung energy in the $\phi$ direction due to the magnetic field. e/$\gamma$ candidates are selected requiring identification criteria using high granularity information from the ECAL and HCAL sub-detectors such as cluster shape and isolation variables.  To distinguish electrons from photons, the presence of a track compatible with the SC is required. The energy and position of the SC is used to propagate a trajectory through the magnetic field to search for compatible hits in the pixel detector. The reconstruction of the electron tracks can be improved by the Gaussian-Sum Filtering (GSF) algorithm which parametrizes the highly non-Gaussian loss of energy of the electron track. The rate of the lowest threshold single electron path ($E_{T}$ threshold = 27 GeV) at different stages of the trigger selection is shown in  Fig.~\ref{fig:ele}a.

Under irradiation the ECAL crystals begin to lose their transparency, which is partly recovered when the radiation stops. This causes the response of the ECAL to vary with time. A laser system monitors the transparency of each crystal and allows corrections to the measured energies to be made.  The effect of irradiation on the efficiency turn-on curves of e/$\gamma$ objects can be seen in Fig.~\ref{fig:ele}b.

\begin{figure}[htb]
\centering
\begin{subfigure}{0.48\textwidth}
\centering
\includegraphics[height=2.3in]{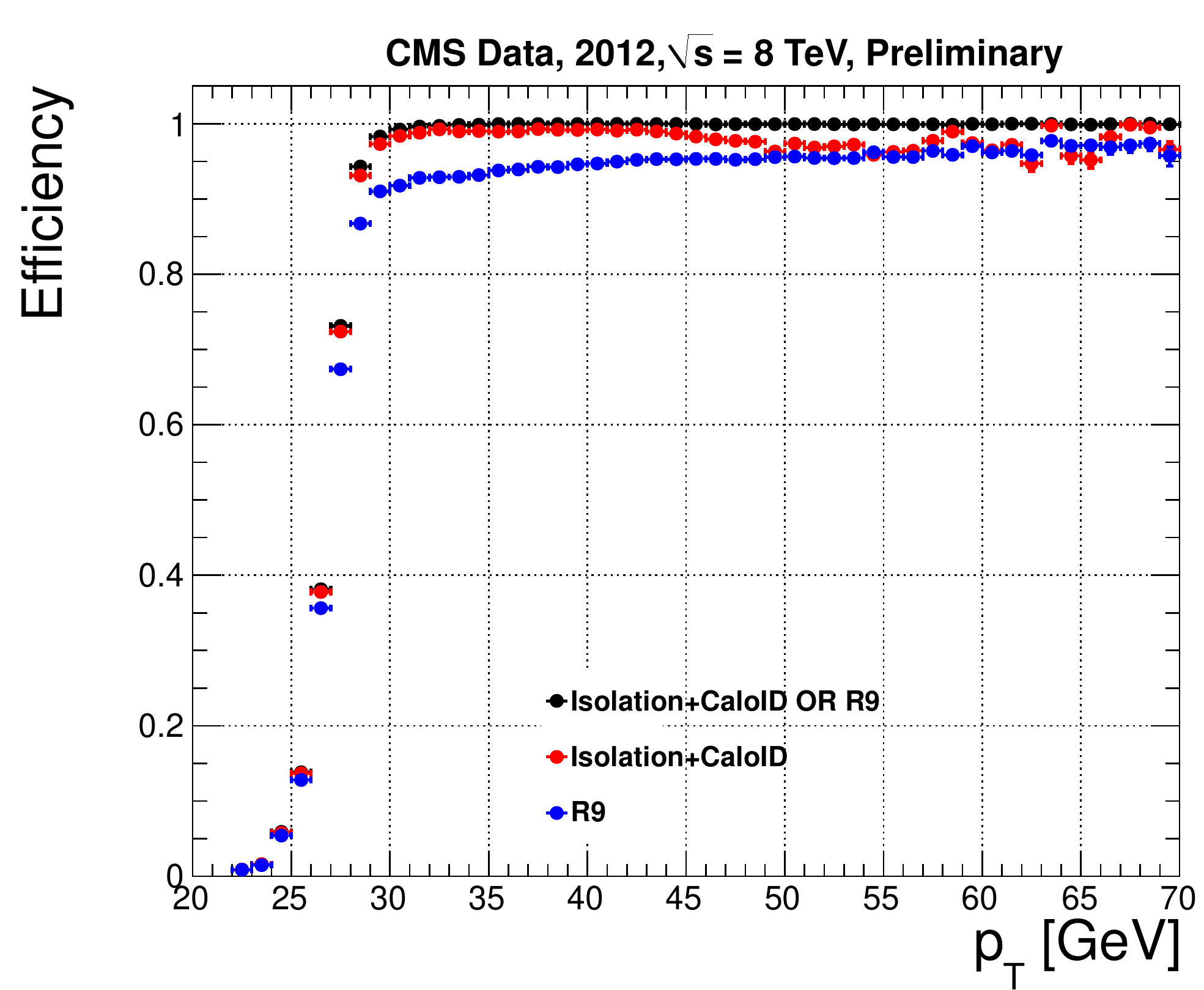}
\caption{}
\end{subfigure}
$\;\;\;\;$
\begin{subfigure}{0.48\textwidth}
\centering
\includegraphics[height=2.3in]{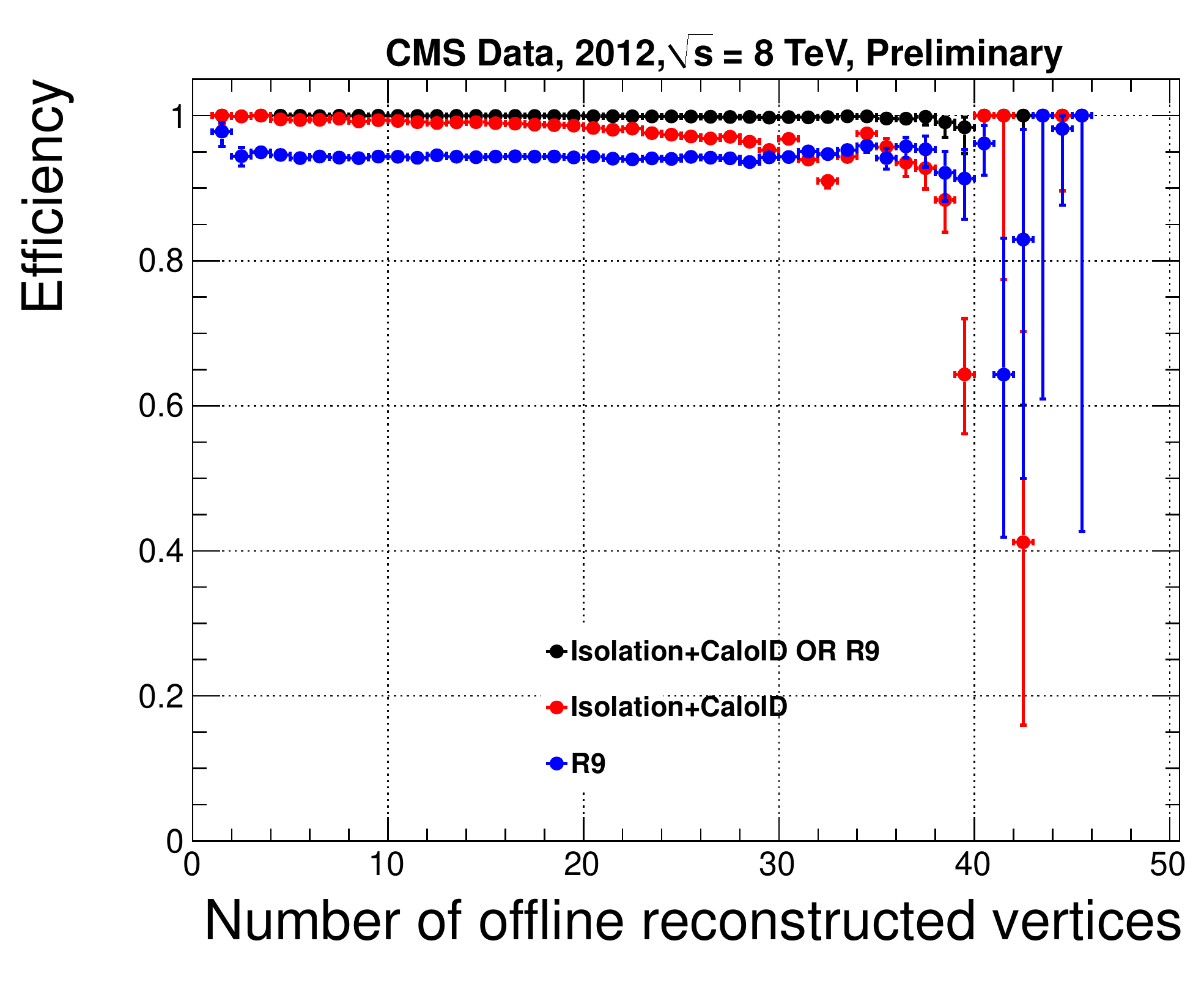}
\caption{}
\end{subfigure}
\caption{Efficiency of the leading leg for double photon trigger as a function of the photon transverse energy and number of vertices comparing with the efficiency of the isolation and calorimeter identification cut as well as the ratio of energy deposited in 3x3 crystals over the energy of the supercluster, referred to as R9.} \label{fig:hlt}
\end{figure}

The efficiency of electron and photon HLT paths is also measured in data using electrons from Z decays.  An example is the efficiency measurement of the double photon trigger used in the H to $\gamma \gamma$ analysis, where the probe electrons are treated as photons i.e the electron track is not considered. The Fig.~\ref{fig:hlt} shows the efficiency curves for this trigger.

\section{Upgrade}

\begin{figure}[htb]
\centering
\begin{subfigure}{0.48\textwidth}
\centering
\includegraphics[height=2.3in]{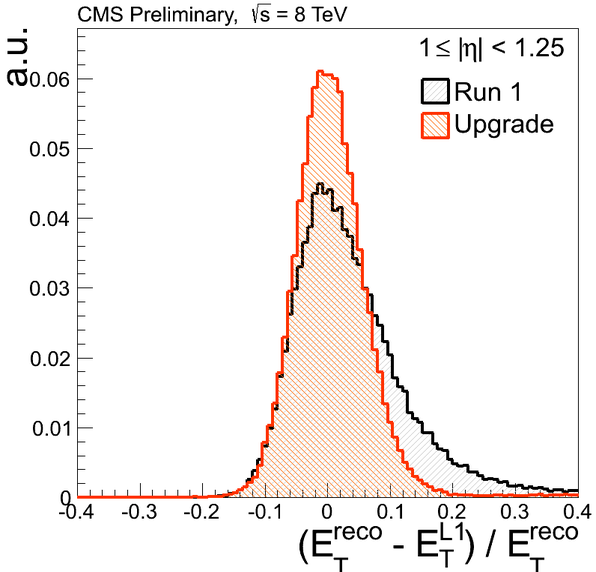}
\caption{The resolution of the upgrade trigger is compared to the current system}
\end{subfigure}
$\;\;\;\;$
\begin{subfigure}{0.48\textwidth}
\centering
\includegraphics[height=2.3in]{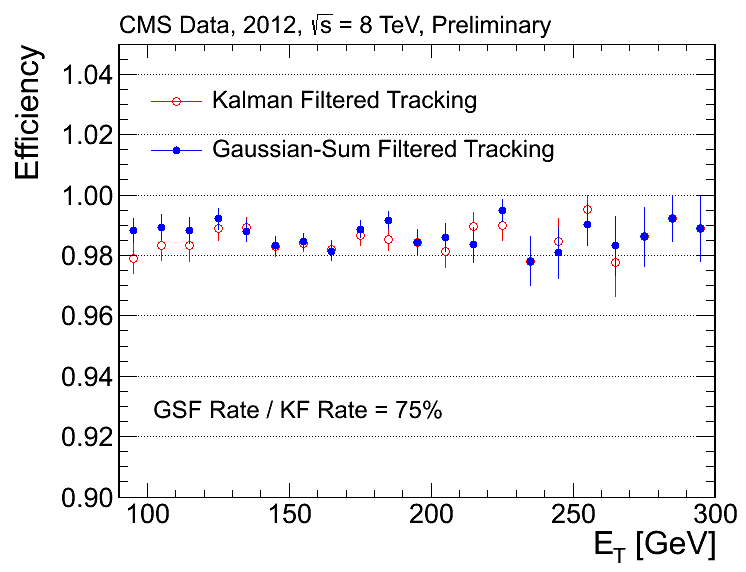}
\caption{The efficiency of identical triggers except they use either Kalman Filtered tracking or GSF tracking}
\end{subfigure}
\caption{} \label{fig:upgrade}
\end{figure}

The upgraded L1 trigger performs a dynamic clustering at the TT level to better aggregate the deposited energy. It results in up to 40\% rate reduction at L1.  The comparison between the current system and the upgraded system clustering resolution is shown in Fig~\ref{fig:upgrade}a.

Through reducing the number of gaussian components used in the GSF refitting, it is possible to reduce the CPU performance without losing efficiency. This allows it to be used as the default tracking algorithm for electrons in the next LHC run. This algorithm results in 25\% rate reduction with no efficiency loss as shown in Fig.~\ref{fig:upgrade}b

\section{Conclusion}

We have described the algorithms used for the online reconstruction of electrons and photons (e/$\gamma$), both at L1 and HLT, and their performance and planned improvements.

\end{document}